# Thermoelectric performance of classical topological insulator nanowires

Johannes Gooth[1*], Jan Göran Gluschke[2,3], Robert Zierold[1], Martin Leijnse[2], Heiner Linke[2], Kornelius Nielsch[1#]

[1] *Institute of Applied Physics, Universität Hamburg, Jungiusstrasse 11, 20355 Hamburg, Germany*

[2] *Solid State Physics and Nanometer Structure Consortium (nmC@LU), Lund University, Box 118, S – 22100 Lund, Sweden*

[3] *School of Physics, The University of New South Wales, Sydney NSW 2052, Australia*

Electronic address: [*] jgooth@physnet.uni-hamburg.de
[#] knielsch@physnet.uni-hamburg.de

ABSTRACT

There is currently substantial effort being invested into creating efficient thermoelectric nanowires based on topological insulator chalcogenide-type materials. A key premise of these efforts is the assumption that the generally good thermoelectric properties that these materials exhibit in bulk form will translate into similarly good or even better thermoelectric performance of the same materials in nanowire form. Here, we calculate thermoelectric performance of topological insulator nanowires based on $Bi_2Te_3$, $Sb_2Te_3$ and $Bi_2Se_3$ as a function of diameter and Fermi level. We show that the thermoelectric performance of topological insulator nanowires does not derive from the properties of the bulk material in a straightforward way. For all investigated systems the competition between surface states and bulk channel causes a significant modification of the thermoelectric transport coefficients if the diameter is reduced into the sub-10 μm range. Key aspects are that the surface and bulk states are optimized at different Fermi levels or have different polarity as well as the high surface to volume ratio of the nanowires. This limits the maximum thermoelectric performance of topological insulator nanowires and thus their application in efficient thermoelectric devices.

## I. INTRODUCTION

Semiconductor nanowires have been predicted to significantly improve upon the corresponding bulk-material's efficiency of thermoelectric (TE) energy conversion,[1-3] quantified by the TE figure of merit $ZT = S^2\sigma T/(\kappa_{el}+\kappa_{ph})$, where $T$ is the temperature, $S$ denotes the Seebeck coefficient, $\sigma$ the electrical conductivity, $\kappa_{el}$ the electron thermal conductivity and $\kappa_{ph}$ the phonon thermal conductivity. While a power factor ($S^2\sigma$) enhancement in nanowires due to quantum dot like states has only been achieved at low temperatures,[4] and the predicted $S^2\sigma$ enhancement due to 1D quantization has yet to be observed,[5-7] it has been shown that utilizing the large surface to volume ratio $s/v$ of nanowires enhancing phonon surface scattering can be used to decreases $\kappa_{ph}$.[8-11] A common approach intended to create high performance TE devices for room temperature operation is therefore to downscale materials with already high bulk $ZT$ to improve the TE performance by taking advantage of the expected reduction of $\kappa_{ph}$ while–this is the rationale of this approach–maintaining the high electronic figure of merit $ZT_{el} = S^2\sigma T/\kappa_{el}$ of the bulk material. In particular chalcogenide nanowires based on $Bi_2Te_3$, $Sb_2Te_3$ and $Bi_2Se_3$ have attracted special interest and inspired extensive research efforts because these compounds are associated with the highest room-temperature bulk TE efficiencies to date, of up to $ZT \approx 1$ for $Bi_2Te_3$.[12] Since $ZT$ strongly depends on the position of the Fermi level of the investigated system, a specific strategy to achieve maximal performance in nano-scaled TE devices is therefore to synthesize nanowires with carrier concentrations optimized for maximum bulk $ZT$.[13-15] However, while the expected reduction in $\kappa_{ph}$ was observed in various thermoelectric transport experiments on such nanowires,[16-19] the observed $S$ was reduced significantly,[19-23] resulting in a substantial decline of $ZT$ ($\approx 0.1$) compared to bulk values. Actually, for intrinsic $Bi_2Te_3$ nanowires (p-type in bulk) *S* was found to be negative (corresponding to n-type transport).[21, 24] Here, we demonstrate that these experimental observations are an expected consequence of the recently discovered topological insulator (TI) nature of such chalcogenide materials.[24-39] Three dimensional TIs are a phase of matter with insulating bulk and with two-dimensional topological surface states that form a single Dirac cone, protected by time reversal symmetry which results in unique transport properties. While topological surface states play no role in TE transport in bulk materials, they may contribute substantially to the transport in nanostructures[40] due to their large surface to volume ratio. In fact, electrical transport experiments on $Bi_2Te_3$[24, 28, 35] and $Bi_2Se_2Te$[41] nanowires in the 200 – 30 nm diameter range revealed that the two-dimensional TI surface channels contribute up to 30-70 % of the total electrical conduction at surface to volume ratios of $s/v = 2 - 5 \cdot 10^{-2}$ nm$^{-1}$. Previous theoretical studies on TE properties of TIs have shown that surface or edge states can in principle enhance the TE performance, for example on two-dimensional TIs with one dimensional edge states,[42, 43] on three-dimensional TIs with line dislocations[44] as well as on $Bi_2Te_3$ and $Bi_2Se_3$ thin films.[45, 46] These calculations focus on length scales below 10 nm, where hybridization of the topological states, resulting in a gap opening at the Dirac point, is expected to enhance the TE efficiency. Such sub-10 nm TE devices have to date not been realized, in part because of limitations in the template size or disturbances of the delicate equilibrium required in self-organized growth processes.[47] Recently, it has been found that in two-dimensional TIs *ZT* is no longer an intrinsic material property but strongly depends on system size just below 10 µm.[48] Due to its relevance for the material science community and the ongoing development of Bi-Sb-Te-Se based TE nanostructures we here consider three-dimensional nanowires with diameters above 10 nm at room-temperature. The thermoelectric performance in topological insulators strongly depends on the effective masses, carrier concentrations, bulk band gap and position of the Dirac point. The Fermi level $E_F$ position is therefore an essential parameter in determining the TE performance of investigated material systems. Our calculations reveal that with decreasing diameter the thermoelectric transport in chalcogenide TI wires is increasingly dominated by the surface states. Because the TE performances of the bulk and of the surface states are optimized at different Fermi level positions, the maximal achievable $S^2\sigma$ and $ZT$ of TI wires with bulk-optimized carrier concentrations can be significantly reduced at diameters in the sub-10 μm range, compared to bulk values. Additionally, the model explains the recently found negative Seebeck coefficient in intrinsic $Bi_2Te_3$ nanowires (p-type in bulk).[21] But even by readjusting the Fermi level for each diameter at the position of maximum TE performance, a significant degradation of the maximum thermoelectric performance in $Bi_2Te_3$ wires for $d < 10$ μm is observed. In contrast, for $Sb_2Te_3$ and $Bi_2Se_3$ nanowires in this diameter range we find a significantly enhanced $ZT$ compared to bulk, mainly because the total thermal conductivity of the wire $\kappa = \kappa_{el} + \kappa_{ph}$ converges to the solely electronic thermal conductivity of the surface states $\kappa_s = \kappa_{el,s}$ with decreasing diameter. Moreover, we show that regardless of precise shape or bulk material, the total thermoelectric efficiency of a topological insulator nanostructure with gapless states will eventually converge to the thermoelectric efficiency of the surface with decreasing system size.

## II. MODEL

We have calculated the room temperature thermoelectric transport coefficients $S$, $\sigma$, $\kappa_{el}$ and $ZT = S^2\sigma T/\kappa$ along the longitudinal axis of a single, cylindrical topological insulator nanowire in the diffusive limit as a function of $E_F$ measured relative to the valence band edge. Standard semi-classical Boltzmann equations under constant relaxation time approximation are used, considering two parallel, non-interacting channels as schematically shown in Fig. 1 (a): A three-dimensional semiconducting bulk channel with two parabolic bands (valence and conduction band) separated by a band gap $\Delta E_b$ and a two-dimensional surface channel with electron and hole cones. The investigated nanowire diameters $d$ lie well above the size range, in which confinement effects in the nanowire bulk[1] or hybridization of the surface states[45, 49] might entail deviations from our calculation for $Bi_2Te_3$, $Sb_2Te_3$ as well as for $Bi_2Se_3$ at 300 K. The detailed methods for calculating $S$, $\sigma$, $\kappa$ and hence $ZT$ as well as all model parameters—obtained from the literature—of the individual transport channels can be found in the supplementary material.[63] For an anisotropic bulk crystal the thermoelectric efficiency $ZT_b$ varies with transport direction, generally determined by the nanowire growth direction, which is expressed in a highly anisotropic effective mass tensor.[63] Here, we choose to perform all calculations along the crystal orientation of highest mobility—parallel to the $a_0$-axis—and therefore along the direction of highest $ZT_b$ to gain maximal bulk contribution in the total thermoelectric transport. Note, that the carrier mobility in nanostructures is generally suppressed compared to bulk values, due to enhanced surface scattering,[50] which could lead to an overestimation of the bulk contribution to the total electrical transport in our calculations. Bulk $Bi_2Te_3$, $Sb_2Te_3$, and $Bi_2Se_3$ are narrow-band-gap semiconductors with band gaps of 105 meV, 90 meV, and 300 meV, respectively. The phonon contributions to the thermal conductivity $\kappa_{ph}$ of the bulk channel were taken from literature and correspond to bulk materials. They do not account for a possible reduction of $\kappa_{ph}$ in the nanowire e.g. due to enhanced surface scattering or impurities. We have chosen to provide both, the electronic part of the thermoelectric figure of merit $ZT_{el} = S^2\sigma T/\kappa_{el}$ as well as $ZT = S^2\sigma T/(\kappa_{el} + \kappa_{ph})$ in order to give an upper and lower limit for the actual TE performance of the nanowire bulk channel.

To capture the transport in the surface states of the nanowire, we consider an energy dispersion $E(k)$ that deviates from an ideal linear dispersion $E(k) = \hbar k v_F$ as observed in angle-resolved photoemission spectroscopy (ARPES).[32, 33] In cylindrical nanowires the energy dispersion relation of the surface states is given by

$$E(k) = \sqrt{\left(\Delta E_{DP} + \hbar k v_F + \frac{\hbar^2 k^2}{2m^*}\right)^2 + \Delta E_s^2} \qquad (1)$$

(Fig. 1 (a)), where $\Delta E_s = 4v_F\hbar d^{-1}$ denotes the energy gap around the Dirac point caused by the periodic boundary conditions around the nanowire with diameter $d$,[51] $v_F$ is the Fermi velocity of the Dirac particles at the Dirac point and $m^*$ is an effective mass term that accounts for the curvature of $E(k)$,[52, 53] leading to an electron-hole asymmetry of the Dirac cone at higher energies.[53] $v_F$ and $m^*$ have been obtained by fitting[61] ARPES measurement data[32,53] for $\Delta E_s = 0$. Note that for $E_F - \Delta E_{DP} > 250$ meV the local curvature of the surface bands $v(k) = \frac{\partial E}{\partial k}/\hbar$ in $Bi_2Te_3$ becomes anisotropic due to hexagonal warping of the Fermi surface,[40] which can cause a asymmetry ratio of 0.4 between the transport directions. For all our calculations we fix $v(k) = v_{a_0}(k)$ and thus $v_{a_0}(0) = v_F$. The position of the Dirac point relative to the bulk valence band edge $\Delta E_{DP}$ has been experimentally determined by ARPES as -265 meV for $Bi_2Te_3$, -38 meV for $Sb_2Te_3$ and 145 meV for $Bi_2Se_3$,[33] exemplary representing materials with the Dirac point deeply buried in a bulk band ($Bi_2Te_3$), at the band edge ($Sb_2Te_3$) and the center of the band gap ($Bi_2Se_3$).

When calculating the total thermoelectric properties of a topological insulator, care must be taken in the qualitative comparison of the surface and bulk contributions because of the different dimensionality of the channels. The total electrical conductivity of a topological insulator is given by

$$\sigma = (G_b + G_s) \cdot \frac{L}{A} = \sigma_b + \sigma_s \frac{s}{v}. \quad (2)$$

$G_b$ and $G_s$ are the electrical conductances of the bulk and the surface channel, respectively; $L$ denotes the length of the nanowire; $A$ its cross-sectional area and $s/v$ its surface-to-volume ratio. Of a (cylindrical) nanowire with diameter $d$ the surface-to-volume ratio accounts to $\frac{s}{v} = \frac{4}{d}$. The total Seebeck coefficient $S$ of a topological insulator nanowire can straightforwardly be calculated within the two-channel model[12] as

$$S = \frac{S_b \sigma_b + S_s \sigma_s \frac{s}{v}}{\sigma_b + \sigma_s \frac{s}{v}} \quad (3)$$

$S_b$ and $S_s$ are the thermopowers of the bulk and the surface channel, respectively. The total electronic part of the thermal conductivity[12] is calculated using

$$\kappa_{el} = \kappa_{el,b} + \kappa_{el,s} \frac{s}{v} + \frac{\sigma_b \cdot \sigma_s \frac{s}{v}}{\sigma_b + \sigma_s \frac{s}{v}} (S_s - S_b)^2 T. \quad (4)$$

$\kappa_{el}$ consists of the individual electronic thermal conductivity of the bulk $\kappa_{el,b}$ and of the surface channel $\kappa_{el,s}$ as well as of an additional diffusion term that is enabled by the two different Seebeck coefficients of the surface states and of the bulk.[12] Note that the total Seebeck coefficient as well as the ratio of the electrical and thermal conductivity of a two channel system will never be larger than the maximum value of the individual channels.[63] The TE efficiency of the bulk or surface states will therefore pose the limit of the TE efficiency of the total nanostructure.

III. RESULTS

The TE transport coefficients of the single bulk channels behave as expected from literature. Their thermoelectric figures of merit $ZT_b(E_F)$ (Fig. 1(b)) as a function of Fermi level reveal the characteristic double-peak structure of a two-band material, resulting from the line shape of the Seebeck coefficient $S_b(E_F)$ (Fig. 1(c)) weighted by the $\sigma/\kappa$-ratios (Fig. 1(d),(e)). The asymmetry in $ZT_b(E_F)$ is caused by the different effective masses of the bulk valence and conductance bands. We obtain maximum bulk thermoelectric figures of merit of $ZT_b(E_{F,opt,b} = 155$ meV$) = 0.84$ for n-type $Bi_2Te_3$, $ZT_b(E_{F,opt,b} = -54$ meV$) = 0.17$ for p-type $Sb_2Te_3$, and $ZT_b(E_{F,opt,b} = 361$ meV$) = 0.06$ for n-type $Bi_2Se_3$ at 300 K for optimal Fermi level positions $E_{F,opt,b}$, corresponding to bulk carrier concentrations of $n_e = 1.11 \cdot 10^{20}$ cm$^{-3}$, $n_p = 1.28 \cdot 10^{19}$ cm$^{-3}$ and $n_e = 4.36 \cdot 10^{18}$ cm$^{-3}$, respectively. These values are in good agreement with literature.[12] Small deviations might occur due to slightly varying bulk parameters in literature. At the surface of thick wires (d > 1 µm) the Dirac cone is quasi-gapless ($\Delta E_s < 1$ meV) and generates two $ZT$ peaks that are shaped by the asymmetric Dirac cones. We obtain maximum surface TE efficiencies of $ZT_s(E_{F,opt,s} = -257$ meV$) = 0.49$ in $Bi_2Te_3$, $ZT_s(E_{F,opt,s} = -42$ meV$) = 0.25$ in $Sb_2Te_3$ and $ZT_s(E_{F,opt,s} = 132$ meV$) = 0.47$ in $Bi_2Se_3$, corresponding to carrier concentrations of about about $n_{e/p} = 2 \cdot 10^{12}$ cm$^{-2}$. This situation covers TIs with gapless surface states in general. For diameters below 1 µm $\Delta E_s$ becomes noticeable in our transport calculations. The maximum $ZT_s$ increases with decreasing nanowire diameter up to $ZT_s(E_{F,opt,s} = -286$ meV$) = 0.91$ in $Bi_2Te_3$, $ZT_s(E_{F,opt,s} = -57$ meV$) = 0.58$ for the hole cone in $Sb_2Te_3$ and $ZT_s(E_{F,opt,s} = 132$ meV$) = 0.87$ for the electron cone in $Bi_2Se_3$ at $d = 10$ nm. We note that the distance between the Fermi level position of the maximum $ZT_s$ and the Dirac point is smaller than $k_BT$. This is a consequence of the asymmetry of the Dirac cone. For an ideal, linear and symmetric cone the optimum $E_F$ would be around $k_BT$ off the Dirac point.

Combining surface and bulk channels, we find that the total thermoelectric properties of topological insulator $Bi_2Te_3$, $Sb_2Te_3$ and $Bi_2Se_3$ wires are increasingly dominated by the surface states with decreasing diameter, as a simple consequence of enhancing surface to volume ratio. If the wire diameter is reduced below the 10 μm range, the competition between surface states and bulk channel causes a significant reshaping of the whole energy dependence of each individual transport coefficient as shown in Fig. 2. While the total electric (Fig. 2 (b),(g),(i)) and thermal conductivity (Fig. 2 (c),(h),(m)) continuously increase as the system-size is decreased, due to the increasing influence of the surface states, the change of the thermopower (Fig. 2 (a),(f),(k)) depends on the Fermi level, increasing or decreasing with declining nanowire diameter. A reduction of the total Seebeck coefficient has been observed in various experiments on Bi-Sb-Te-Se based nanostructures such as nanowires[19-23] and thin films.[54-56] For $Bi_2Te_3$ the transformation of the energy landscape has an additional crucial impact: Below a nanowire diameter of about 200 nm, the thermopower becomes of negative sign for Fermi levels within the bulk valence band, because the Dirac point on the surface of $Bi_2Te_3$ is located in the bulk valence band leading to a bipolar competition between electrons on the surface (negative

Seebeck coeffcicient) and holes in the bulk (positive Seebeck coefficient). This finding possibly explains recent thermal[21] and magnetotransport measurements[24] on $Bi_2Te_3$ nanowires, wherein a negative thermopower has been determined at Fermi levels deeply buried in the bulk valence band. The relative contribution of the surface states to the total electrical conductivity ($Bi_2Te_3$: 15% for $d = 100$ nm at $E_F = -205$ meV) are in reasonably good agreement with experimental data.[24, 28, 35] The herein presented calculations are done for single-crystalline nanostructures with perfect surfaces and there are several possible reasons for why the relative contributions of the electronic thermal and electric conductivity could deviate in experiments: On the one hand, imperfections at the surface of the nanowire could lead to a lower mean free path than assumed in our model or different dominating scattering mechanisms might entail longer relaxation times in the surface channel.[48] On the other hand, additional scattering mechanisms in the bulk channel due to grain boundaries or surface roughness are known to decrease its electrical and thermal conductivity.[8-11] Because all listed effects would result in an overall reduction of the total electrical and thermal conductivity of the topological insulator nanowire, an increasing electrical and thermal conductivity with decreasing nanowire diameter seems to be a strong indication for topological surface states.

We now discuss the common experimental strategy to obtain the predicted high thermoelectric performances in nanowires by downscaling bulk material with carrier concentrations optimized for highest bulk $ZT_b$[13-15] in order to retain the high $ZT_{el}$ and benefit from the suppression of $k_{ph}$. In our calculation this experimental strategy is similar to a reduction of the nanowire diameter, fixing the Fermi level at the value that yields the highest bulk $ZT$ $E_{F,opt,b}$.[63] At 10 nm nanowire diameter we obtain $ZT_{el}(E_{F,opt,b} = -155$ meV)

= 0.01 for a n-type $Bi_2Te_3$, $ZT_{el}(E_{F,opt,b} = -54$ meV$) = 0.42$ for a p-type $Sb_2Te_3$ and $ZT_{el}(E_{F,opt,b} = 361$ meV$) = 0.04$ for a n-type $Bi_2Se_3$. Analogous system-size dependencies are observed in the case of full phonon contribution. $ZT(E_{F,opt,b})$ converges to similar values as $ZT_{el}(E_{F,opt,b})$ at 10 nm diameter. Because the increasing electrical conductivity is compensated by the likewise increase of the thermal conductivity, the thermoelectric efficiency is dominated by $S$. The results reflect the overall experimental observations very well. A more detailed comparison of theory and experiment requires diameter dependent thermoelectric transport studies on nanowires at fixed Fermi levels. Such a study is available for $S$ and $\sigma$ of $Sb_2Te_3$ nanowires,[57] where an increase in $S$ from $S < 90$ μV/K at $d = 100$ nm to $S \approx 110$ μV/K at $d = 20$ nm was observed, which compares well with our findings (see Fig. 3 in the supplementary material[63]). Our results show that such an increase can be caused by the increasing influence of the surface states, even when size quantization remains irrelevant. While $ZT_{el}(E_{F,opt,b})$ and $ZT(E_{F,opt,b})$ for $Bi_2Te_3$ and $Bi_2Se_3$ nanowires are supressed, because the TE performance of the nanowire bulk and surface are optimized at different Fermi levels, $ZT_{el}(E_{F,opt,b})$ and $ZT(E_{F,opt,b})$ of $Sb_2Te_3$ is slightly enhanced compared to $ZT_b(E_{F,opt,b})$. However, in comparison to their electronic bulk counterparts, $ZT_{el}(E_{F,opt,b})$ of wires with diameters in the nanometer range is strongly suppressed for all materials investigated.

The question thus arises whether topological insulator nanowires can reach bulk $ZT_{el}$ at all to benefit from the $\kappa_{ph}$ suppression for improving the total TE performance? The remarkable suppression of the electronic thermoelectric efficiency in nanometer-scaled optimized bulk materials can be attributed to two main features: first, to a narrowing of the $ZT$-peaks as the system size is reduced; and second, to a Fermi level-shift of the efficiency maximum. Both effects warrant a very accurate repositioning of the Fermi level at each nanowire diameter to maximize the thermoelectric performance at the specific system size. Tuning the Fermi level (i.e. the carrier concentration) via field effect or doping has recently turned out to provide a powerful experimental tool to achieve maximum thermoelectric performance in semiconducting nanostructures.[4, 58, 59] Following this route, we extract the maximum $ZT$ and $ZT_{el}$ at the optimized Fermi level positions $E_{F,opt}$ for each nanowire diameter as shown in Fig. 3 (a) and (b). For all materials investigated, a non-monotonic relation between TE efficiency and system size is obtained. A minimum splits the $ZT$ and $ZT_{el}$ curves into two distinct size ranges marking the crossover between surface state dominated transport in smaller systems and bulk dominated transport in larger systems. The $ZT_{el}$ curves are shaped by the thermopower at $E_{F,opt}$ evolving from competing $S_S$ and $S_b$. The dependence of $ZT$ on the system size follows the same arguments as $ZT_{el}$, but the bulk channel is additionally weighted by $\kappa_{ph}$. $\kappa_{ph}$ results in a suppression of $ZT_b$ as well as in an overall suppression of the total nanowire efficiencies $ZT$ compared to $ZT_{el}$. This demonstrates that phonons still play a noticeable role in thermoelectric transport, even if topological surface states dominate the electronic transport properties of the nanowire.

We find that nanoscaling $Bi_2Te_3$ wires leads to a decrease of the maximum $ZT$ compared to bulk (by a factor three at $d = 10$ nm). In contrast, the maximum $ZT$ for $Sb_2Te_3$ and $Bi_2Se_3$ is increased (by a factor of ten [three] for $Sb_2Te_3$ [$Bi_2Se_3$] at $d = 10$ nm). In all nanowire systems investigated the maximum $ZT_{el}$ is at least one order of magnitude suppressed compared to their bulk counterparts (at $d = 10$ nm), because the total maximum TE performances of the topological insulator nanowires converge towards the maximum TE performance of the surface states with decreasing diameter (the increase in $ZT$ for $Sb_2Te_3$ and $Bi_2Se_3$ at small diameters occurs mainly because the contribution of phonon thermal conductivity in the nanowire bulk becomes small compared to the solely electronic thermal conductivity contribution of the surface states, not because the surface states have a very high $ZT_{el}$). Similar findings are obtained for topological insulators with gapless surface states. Exemplary calculations for gapless Dirac cone systems on thin films are shown in Fig. 3 (d) and (e).

## IV. DISCUSSION

Summing up, in the presence of surface states, the total thermoelectric efficiency of a topological insulator nanostructure will eventually converge to the thermoelectric efficiency of the surface states with decreasing system size, regardless of precise shape or bulk material. However, the details of this process depend on the surface to volume ratio of the nanostructure, on $ZT_b$, on the relaxation time of the surface states as well as on the relative position of the Dirac point to the bulk band structure. Topological insulator nanostructures could therefore, on one hand be a chance to enhance the thermoelectric performance of materials with $ZT_b < ZT_s$ and on the other hand pose a limit to the thermoelectric performance of materials with $ZT_b > ZT_s$. Nevertheless, the maximum TE surface efficiencies $ZT_s \approx 0.5$ ($Bi_2Te_3$) of gapless surface states and $ZT_s \approx 0.9$ ($Bi_2Te_3$) on the surface of cylindrical nanowires (at 10 nm diameter) do not exceed the highest room-temperature bulk TE efficiencies to date and are far below the electronic room-temperature bulk TE efficiencies. A reduction of $\kappa_{ph}$ does not lead to an enhancement of the TE efficiency of topological insulator nanostructures beyond best bulk efficiencies ($ZT \approx 1$) to date. Important for future room temperature TE applications will therefore be the suppression of the topological surface states. The implementation of magnetic impurities on the surface of a topological insulator is known to break time reversal symmetry and thus to introduce a backscattering channel accompanied with a gap opening at the Dirac point.[60, 61] In addition, recent calculations predict that the hybridization gap formed in ultrathin topological insulators with system sizes below 10 nm may provide a chance to improve their thermoelectric properties far beyond $ZT = 1$.[42-46] Topological insulator nanowires with diameters below 10 nm could hence be of high interest because hybridization of the surface states as well as quantum confinement in the bulk is proposed to significantly enhance the thermoelectric performance of each individual transport channel. However, in the diameter range experimentally achievable to date ($d > 10$ nm), the presence of surface states limits the thermoelectric efficiency of topological insulator nanowires and thus their application in efficient thermoelectric devices. In this size range, alternative concepts utilizing confinement effects in nanometer-scaled non-conventional good bulk thermoelectric materials, such as in InAs,[4] ZnO and GaN[62] nanowires as well as in Ge-Si core-shell nanowires[58] might pave the way to future TE application.

**ACKNOLEDGEMENTS**

We thank Ulrich Merkt and Toru Matsuyama for useful discussions. This work was supported by the Deutsche Forschungsgemeinschaft (DFG) via Graduiertenkolleg 1286 "Functional Metal-Semiconductor Hybrid Systems", Project NI-616/18-1 and SPP 1666 "Topological insulators: Materials – Fundamental Properties – Devices", by the Swedish Energy Agency (project 38331-19) as well as by the Swedish Foundation for Strategic Research.

**FIGURES**

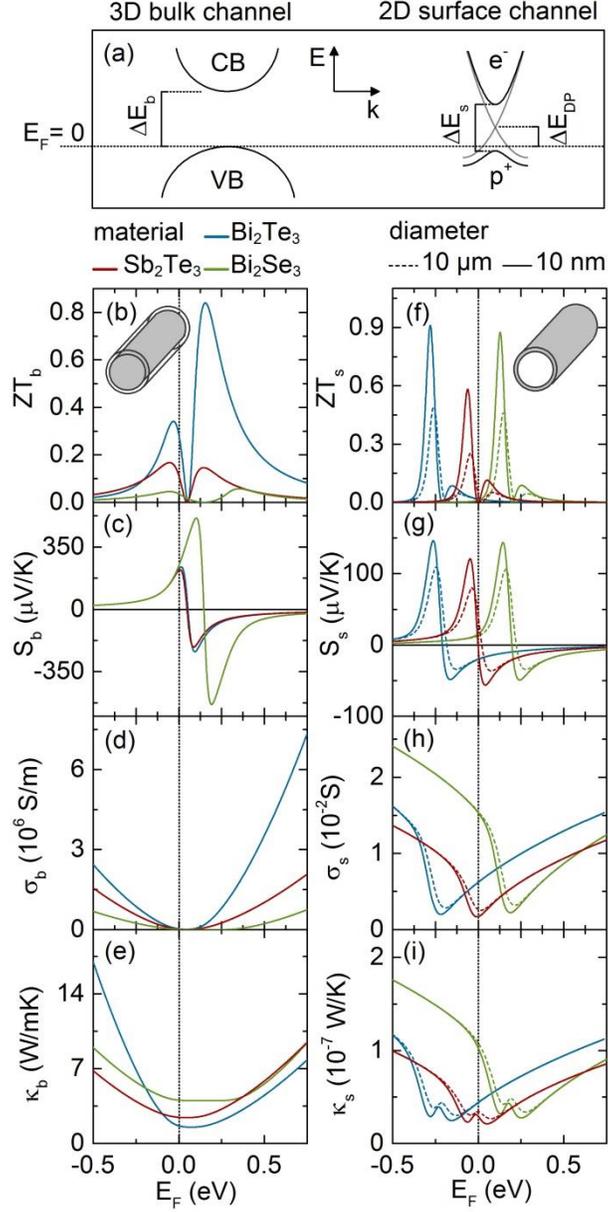

**FIG 1.** Thermoelectric transport coefficients of the bulk (left) and of the surface (right) channel of cylindrical topological insulator $Bi_2Te_3$ (blue), $Sb_2Te_3$ (red) and $Bi_2Se_3$ (green). Bulk and surface channel correspond to the subscripts b and s respectively. (a) The three-dimensional bulk channel is a two-band semiconductor with parabolic dispersion relation, where the two bands (valence (VB) and conduction band (CB)) are separated by a bandgap $\Delta E_\text{b}$. The transport parameters of the bulk are independent of the nanowire diameter $d$. The two-dimensional surface channel is characterized by a Dirac cone with electron states (e⁻) on one side of the Dirac point and hole states (p⁺) on the other. Angular momentum states around the nanowire perimeter cause a gap $\Delta E_\text{s} \approx 4 v_\text{F} \hbar d^{-1}$ around the Dirac point. $\Delta E_\text{DP}$ is the distance between the Dirac point and the bulk valence band edge. (b), (f) The thermoelectric figure of merit $ZT$, (c), (g) the thermopower $S$ as well as the (d), (h) electrical $\sigma$ and (e), (i) thermal conductivity $\kappa$ of both channels are calculated as a function of the Fermi level $E_\text{F}$, measured relative to the bulk valence band edge.

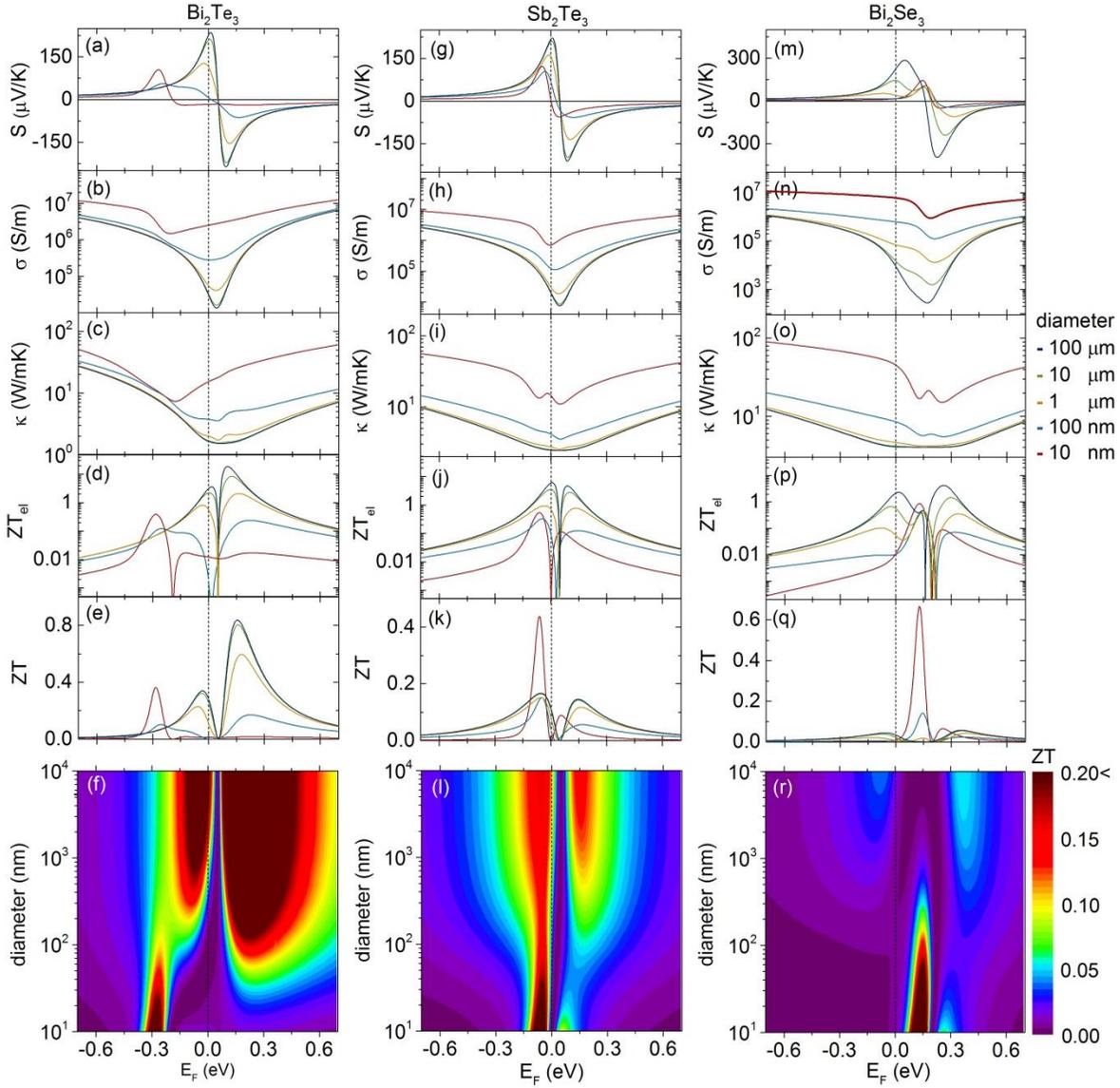

**FIG 2.** Diameter-dependent thermoelectric transport coefficients of topological insulator Bi$_2$Te$_3$, Sb$_2$Te$_3$ and Bi$_2$Se$_3$ nanowires (from left to right column). (a), (g), (m) thermopower $S$; (b),(h),(n) electrical conductivity $\sigma$; (c), (i), (o) thermal conductivity $\kappa$; (d), (j), (p) electronic figure of merit $ZT_{el}$ and (e), (k), (q) thermoelectric figure of merit with full phonon contribution $ZT$ at 300 K are plotted as a function of Fermi level, relative to the bulk valence band edge ($E_F = 0$ eV). (f), (l), (r) $ZT$ is plotted as function of diameter and Fermi level $E_F$ in a two-dimensional color plot. When the wire diameter is reduced the competition between surface states and bulk channel causes a significant reshaping of the whole energy dependence of each individual transport coefficient, because nanowire bulk and surface have different Fermi level dependencies and with decreasing wire diameter the thermoelectric transport is increasingly dominated by the surface states as a simple consequence of increasing surface to volume ratio.

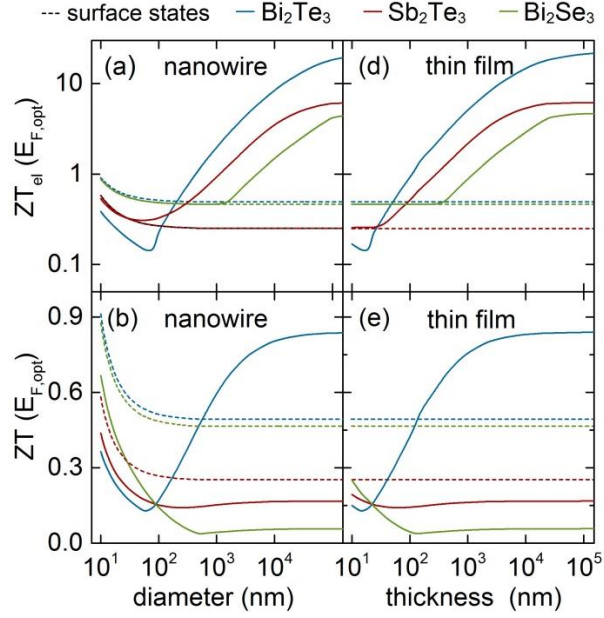

**FIG 3.** Maximum electronic thermoelectric figure of merit $ZT_{el}(E_{F,opt}) = S^2\sigma/\kappa_{el}$ and maximum thermoelectric figure of merit $ZT(E_{F,opt}) = S^2\sigma/(\kappa_{el}+\kappa_{ph})$ at optimized Fermi level position $E_{F,opt}$ of a topological insulator $Bi_2Te_3$ (blue), $Sb_2Te_3$ (red) and $Bi_2Se_3$ (green) nanowire (left) and thin film (right) as a function of diameter and thickness, respectively. The dotted lines show the maximum $ZT$ of the single surface states. Regardless of precise shape or bulk material, the total thermoelectric efficiency of a topological insulator nanostructure with gapless surface states will eventually converge to the thermoelectric efficiency of the surface with decreasing system size.